# Parallel magnetic field induced strong negative magnetoresistance in a wide p-Ge$_{1-x}$Si$_x$/Ge/p-Ge$_{1-x}$Si$_x$ quantum well

*M. V. Yakunin*[*], G. A. Alshanskii[*], Yu. G. Arapov[*], V. N. Neverov[*]
and  O. A. Kuznetsov[#]

[*] Institute of Metal Physics RAS Ural branch, Ekaterinburg GSP-170, 620219 Russia
[#] Scientific Research Institute at Nizhnii Novgorod State University, Russia
e-mail: yakunin@imp.uran.ru

**Abstract.** A negative magnetoresistance (NMR) under the in-plane magnetic field, reaching maximum 30-40% of its zero-field value in fields ≥12 T, has been found in wide Ge$_{1-x}$Si$_x$/Ge/p-Ge$_{1-x}$Si$_x$ quantum wells (QW) containing the quasi-two-dimensional hole gas. In the QWs of intermediate widths and hole densities, this NMR may be explained as being caused by suppression of the intersubband scattering due to the upper subband depopulation. For the widest QWs with the highest hole densities, in which the hole gas is divided into two sublayers, the NMR is tentatively interpreted as also been due to suppression of the intersubband scattering, but subbands are the lowest symmetric and antisymmetric states of the double quantum well structure. These subbands shift under the in-plane magnetic field not vertically in energy, but horizontally along the wave vector.

In this paper we present the results of our further study of magnetotransport due to holes in a Ge$_{1-x}$Si$_x$/Ge heterosystem [1], now with the focus on phenomena under the in-plane magnetic fields.

The samples were Ge$_{1-x}$Si$_x$/Ge multy-layers, differed by the Ge quantum well (QW) width $d_w$ and the hole densities per Ge layer $p_s$. Concrete parameters are given in figures 1*a-c* together with the experimental results. The Ge$_{1-x}$Si$_x$ barriers, sufficiently wide to prevent the inter-Ge-layer tunneling, were selectively doped with boron in their central parts. Low temperature hole mobilities amounted 1-1.4 m$^2$/V·s.

As has been found from the quantum Hall effect (QHE) [1], the hole gas is integrated through the whole Ge layer width (i.e. may be treated as a single quasi-2D hole gas) in structures 1123 and 1125, but is divided into two 2D sublayers in structures 475 and 476, having (nominally equal) the widest wells and highest hole densities. This separation is caused by the well bottom bending due to hole-hole repulsion and selective doping that forces the holes to locate at the opposite interfaces of the Ge layer (see the left insets in fig.1). A brilliant indication of the difference between the integrated and divided hole gases is either existence or absence of the QH plateau for the filling factor $v = 1$. In samples 1123 and 1125 these plateau pronouncedly existed concomitantly with a deep minimum in the longitudinal magnetoresistance, while the $v = 1$ peculiarities were missed in samples 475/476.

As is seen in fig.1, the common experimental feature in the parallel magnetic fields for all the samples investigated is a negative magnetoresistance (NMR) reaching 30-40 % of the zero field value in fields $B_\parallel \geq \sim 12$ T. In samples 1123 and 1125 this NMR should be due to the well-known effect of the diamagnetic shift of the upper electric subband that results in its depopulation and suppression of the intersubband scattering, initially present. The NMR may reach a value up to 50 % for this mechanism [2].

For sample 1123 the magnetoresistance $\rho$ is nearly independent of $B_\parallel < \sim 6$ T (fig.1*a*), with subsequent fast drop reaching the maximum slope near $B = 10$ T. For sample 1125 (fig.1*b*) the drop of $\rho(B_\parallel)$ starts from the very beginning and reaches the maximum slope at $\sim 6.5$ T. The difference between the two samples correlates with the data in perpendicular fields [3], where a radical difference in the structure of experimental recordings for the longitudinal and Hall magnetoresistances has been found for these two samples and explained as that, while the Fermi level lies at certain depth inside the upper subband of sample 1123, it is in close vicinity of the upper subband edge for sample 1125. A degree of the upper subband involvement may be roughly estimated as being proportional to $p_s d_w^2 = 1.9$ and 0.9 for samples 1123 and 1125 respectively [3]. In parallel fields the point of maximum slope in $\rho(B_\parallel)$ is usually ascribed to the moment when the Fermi level leaves the upper subband. Then, the more involved upper subband in sample 1123 manifests in the higher field of its depopulation, in agreement with the results in perpendicular fields. Our calculations of the energy levels verify these experimental findings (see the insets in figures 1*a,b*).

Surprisingly, the $\rho(B_\parallel)$ curves for samples 475/476 have turned out to be similar to those for sample 1125: the NMR that starts from the weakest fields and reaches the amplitude of ~ 30 % the zero field value at fields above 10 T (fig.1*c*). It is in spite of quite different layout of the energy levels in these samples (inset in fig.1*c*): the lowest levels $E_0$ and $E_1$ almost coincide, in accordance with nearly a complete separation of the hole gas into sublayers, and the Fermi level lies much higher of them both. Thus, the simple picture of some level depopulation due to diamagnetic shift is impossible in this case.

Then, what is the cause of a pronounced NMR in samples 475/476? The NMR may be due to suppression of quantum corrections to the conductivity. These would manifest for the case $k_F l \gg 1$ ($k_F$ – the Fermi wave vector, $l$ – a mean free path) and have the value ~ $1/k_F l$. For samples 457/476: $k_F l = 10 \div 15$, so the maximum effect due to the quantum corrections should be less than 10 %. These corrections could be responsible for the NMR in relatively low fields, but cannot yield the observed NMR of about 30 % in the high fields. All the other mechanisms known should lead to a positive magnetoresistance: MR in the vicinity of the metal-insulator transition [4]; MR due to strong coupling of the parallel field to the orbital motion in the quasi-2D systems [5]; due to interplay between spin polarization and the screening effects [6] etc.

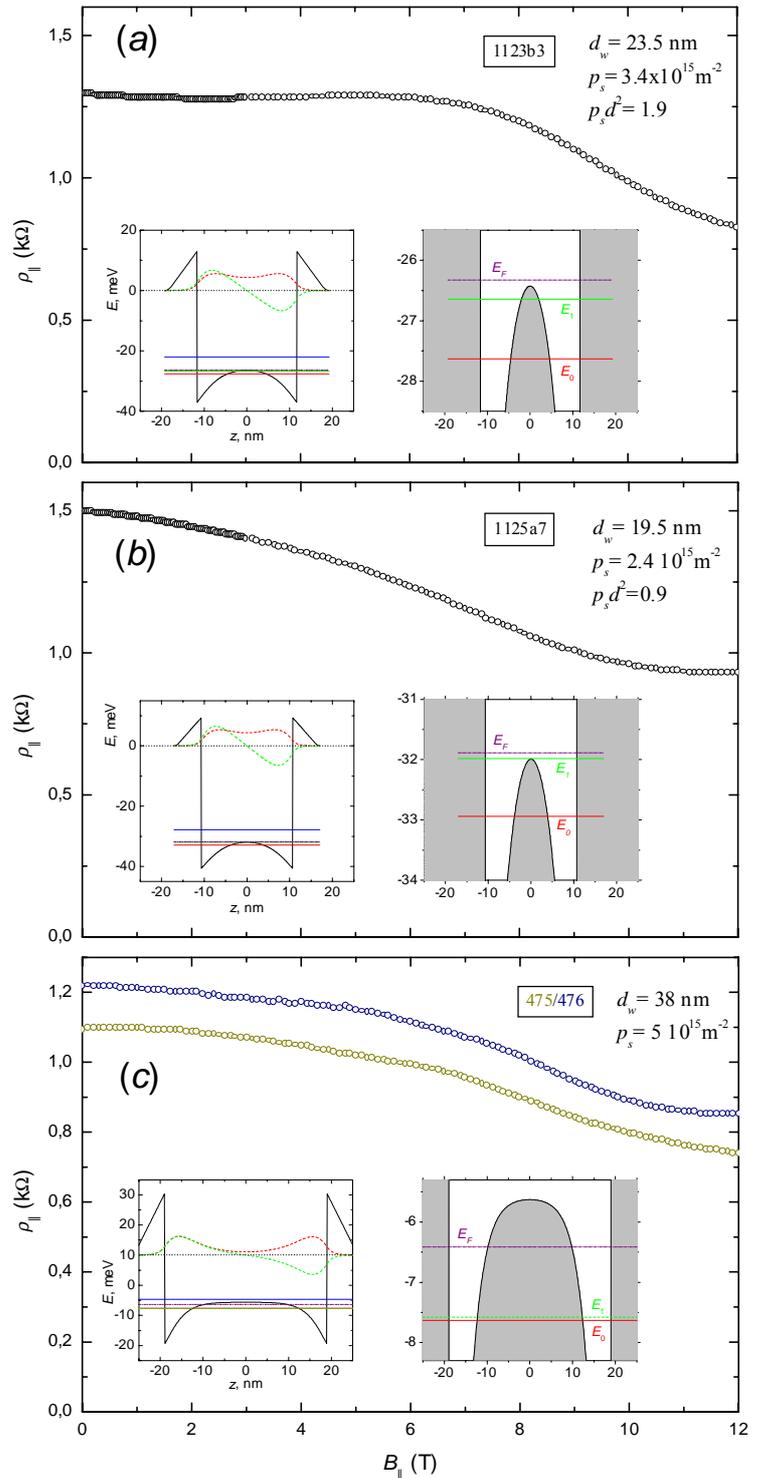

**Fig.1.** Magnetoresistance in the parallel magnetic fields of samples 1123 (*a*), 1125 (*b*) and 475/476 (*c*).
In the insets: calculated energy profiles, energy levels, Fermi level and the wave functions.

By these reasons we suppose that nature of the NMR in samples 475/476 also originates from the suppression of an intersubband scattering, but the mechanism of this scattering is unlike that in the samples 1123 and 1125. The essential difference stems from that, while in samples 1123 and 1125 the potential in the Ge layer may be treated as a single QW, this of the samples 475/476 is effectively a double quantum well (DQW).

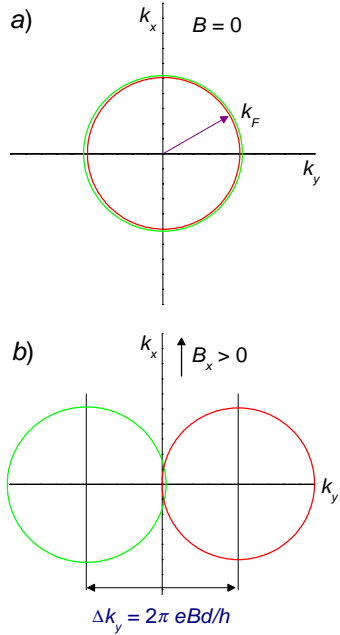

**Fig.2**. Fermi contours in a $k_xk_y$ plane: coincided in zero field (*a*) and shifted along $k_y$ in the field $B = B_x$ (*b*).

In a DQW, a parallel magnetic field $B = B_x$ shifts the Fermi contours of the QWs in the plane direction along $k_y$ on the value $\Delta k_y = eBd/\hbar$, with $d$ – an effective distance between gases in the sublayers [7] (fig.2). Whilst in a nearly symmetrical DQW in zero magnetic field some finite probability exists for transitions between the two coinciding Fermi circles (fig.2*a*), it drops down as the circles shift in *k*-space relative each other under parallel field (fig.2*b*). This drop has been directly observed in the magnetotunneling experiments on a traditional DQW [8]. Whereas in a single QW the NMR due to depopulation of the upper subband stemmed from the energy conservation law, the suppression of the intersubband scattering in a DQW would stem from the momentum conservation law. Also, a parallel field acts to increase the separation between the sublayers [9] that would also weaken the transitions between them.

Amazingly, in spite of a considerable number of existing experimental investigations of $\rho(B_{\parallel})$ in a DQW, attention has not been paid yet to the NMR. Maybe this is due to the fact that in the GaAs/n-AlGaAs based DQW (where almost all the experiments in DQW have been performed) a small electron mass causes considerable anticrossings of the $E(k_{//})$ paraboloids, which manifests in the existence of sharp structures in $\rho(B_{\parallel})$ [10-12]. A considerable, up to ~30 %, NMR presents in some experimental results [11,12], albeit without explanation. Notably, in a paper [11] experiments have been performed in the self-formed DQWs inside the wide QWs, as in our case, but containing the electron gas. The large hole mass in our experiments creates new conditions with very small anticrossings that lead to the absence of the sharp structures in $\rho(B_{\parallel})$ and probably – to the more pronounced monotonous NMR.

The drop of $\rho(B_{\parallel})$ due to the interlayer scattering should last in $B_{\parallel}$ until the two Fermi circles leave each other at $\Delta k_y \approx 2k_F$ (fig.2*b*). Estimations for samples 475/476 yield the value of relevant field $B^* \approx 5$ T for $d = 33$ nm taken as the distance between the wave functions maxima (fig.1*c*). Agreement with the experimentally observed drop of $\rho(B_{\parallel})$ till $B^* \geq 10$ T may be achieved with about a factor of two smaller effective distances. The explanation of this controversy may be in that, although the wave function in the central region of the well is weaker, the intersublayer scattering probability is exponentially stronger here due to smaller distances. Also, the finite width of the heterojunctions would result in the smaller effective QW width leading to the reduced intersublayer distance.

*Acknowledgement*

The work is supported by RFBR, projects 99-02-16256 and 01-02-17685.